\documentclass[fleqn,12pt]{wlscirep}
\usepackage{amsmath}
\usepackage{graphicx}
\usepackage{footmisc}
\usepackage{color}

\begin{document}

\title {Electron liquid state in the symmetric Anderson lattice}

\author[1]{Igor N. Karnaukhov}
\affil[1]{G.V. Kurdyumov Institute for Metal Physics, 36 Vernadsky Boulevard, 03142 Kiev, Ukraine}
\affil[*]{karnaui@yahoo.com}

\begin{abstract}
Using mean field approach, we provide analytical and numerical solution of the symmetric Anderson lattice for arbitrary dimension at half filling.
The symmetric Anderson lattice is equivalent to the Kondo lattice, which makes it possible to study the behavior of an electron liquid in the Kondo lattice.  We have shown that, due to hybridization (through an effective field due to localized electrons) of electrons with different spins and momenta $\textbf{k}$  and $\textbf{k}+\overrightarrow{\pi}$,  the gap in the electron spectrum  opens at half filling. Such hybridization breaks the conservation of the total magnetic momentum of electrons, the spontaneous symmetry is broken. The state of electron liquid is characterized by a large Fermi surface. A gap in the spectrum is calculated depending on the magnitude of the on-site Coulomb repulsion and value of s-d hybridization for the chain, as well as for square and cubic lattices. Anomalous behavior of the heat capacity at low temperatures in the gapped state, which is realized in the symmetric Anderson lattice, was also found.
\end{abstract}
\maketitle

\section*{Introduction}
The processes of electron scattering with spin flip play a dominant role in the Kondo problem, they are determined by the exchange s-d interaction \cite {W1, W2}. They should also be explicitly taken into account when considering the behavior of the Kondo lattice, since at half-filling the insulator state is also determined by this interaction. The one-particle modifications of the Kondo lattice (see for example \cite{1}), exact solvable many-particle models \cite{K0,KS}, which do not taken into account the electron scattering with spin flip unable to describe the Kondo insulator state. Numerical calculations of the Anderson and Kondo lattices take into account clusters with a small number of particles,  which significantly affects the adequacy of the results obtained \cite{A1,A2,A3}. This leads to the fact that the mechanism of the formation of the large Fermi surface and the gap in the electronic spectrum in the Kondo lattice at half filling is still unclear. In contrast to the single-impurity Anderson and Kondo models in their lattice version, the processes of electron scattering with spin flip cannot be calculated exactly.

In the weak s-d hybridization limit the Hamiltonian of the Anderson lattice reduces to that of the Kondo lattice. It should also be taken into account, that the local density of localized electrons must be equal to unity. In the weak s-d hybridization limit, the local density of d-electrons in the symmetric Anderson lattice is equal to unity.  d-electrons determine the local spin-$\frac{1}{2}$ at the sites of the lattice,  only in this case we can talk about the spin-$\frac{1}{2}$ Kondo lattice. Thus the symmetric Anderson model is similar to the Kondo lattice.
In the strong coupling limit of the Hubbard model, when $U \gg t$ ( $U$ and $t$ are on-site repulsion and hopping integral),
it is shown that the Hubbard model and the Kondo lattice model become identical \cite {3}. It gives possibility to use the formalism, proposed for calculation of the Mott transition in the Hubbard model \cite{K1}, for solution of the Kondo lattice problem.

In the paper, we consider the solution of the symmetric Anderson model, using a mean field approach. An effective $\lambda$-field connects the states of d-electrons with the different spins and momenta $\textbf{k}$, $\textbf{k}+\overrightarrow{\pi}$. Due to the s-d hybridization, the states of s-electrons
with different spins and momenta $\textbf{k}$, $\textbf{k}+\overrightarrow{\pi}$ also hybridize, the gap opens in the electron spectrum at half filling. The value of the gap is determined by the magnitude of the $\lambda$-field (which in turn depends on the on-site repulsion) and the value of the s-d hybridization.
We shall show that breaking spontaneous symmetry in the Anderson lattice Hamiltonian makes it possible to take into account the scattering of conduction electrons on d-electrons with a spin flip. They determine the ground state of symmetric Anderson and Kondo lattices, the value of the gap in the electron spectrum, cause formation of  an insulator state at half-filling. The electron spectrum is symmetric about the zero energy, it corresponds to the symmetric Anderson model for a nontrivial solution of the $\lambda$-field. The gap in the electron spectrum has a many-particle nature.

\section*{Model}
The Hamiltonian of the Anderson lattice is the sum of two terms, the first of which is determined by energy of the bands of s- and d-electrons and hybridization between them, the second takes into account the on-site repulsion of d-electrons ${\cal H}={\cal H}_{0}+{\cal H}_{int}$
\begin{eqnarray}
&&{\cal H}_0= -
\sum_{j=1}^{N-1}\sum_{\sigma=\uparrow,\downarrow}
(c^\dagger_{j,\sigma} c_{j+1,\sigma}+ c^\dagger_{j+1,\sigma} c_{j,\sigma})
+
\epsilon_g \sum_{j=1}^{N}\sum_{\sigma=\uparrow,\downarrow}n_{j,\sigma}+v\sum_{j=1}^{N}\sum_{\sigma=\uparrow,\downarrow}
(c^\dagger_{j,\sigma} d_{j,\sigma}+ d^\dagger_{j,\sigma} c_{j,\sigma}),\nonumber\\&&
{\cal H}_{int}= U\sum_{j=1}^{N}n_{j,\uparrow}n_{j,\downarrow},
\label{eq:H}
\end{eqnarray}
where $c^\dagger_{j,\sigma},c_{j,\sigma}$ and $d^\dagger_{j,\sigma},d_{j,\sigma}$ ($\sigma=\uparrow,\downarrow)$ are the fermion operators determined on a lattice site $j$, $U$ is the  value of the on-site Hubbard interaction determined by the density operator $n_{j,\sigma}=d^\dagger_{j,\sigma}d_{j,\sigma}$, the band width of c-fermions is determined by the hopping integral equal to one, the energy of flat band of d-fermions equal to $\epsilon_g$,  $v$ defines the hybridizations of s- and d-electrons, N is the total number of atoms.

First of all, we will focus on the consideration of the model for single Anderson impurity, exact solution of which has been obtained by  Wiegmann \cite{W1}. At $\epsilon_g=-\frac{U}{2}$ and $U \gg \Gamma $ ($\Gamma= v^2$) trickly one electron is localized on the impurity, so we can talk about an impurity with spin-$\frac{1}{2}$. This case corresponds to the symmetric Anderson model. According to \cite{W2} the behavior of impurity in the symmetric Anderson  model is equivalent to the spin-$\frac{1}{2}$ Kondo impurity. In this context, in a non-magnetic (paramagnetic) state the behavior of an electron liquid in the symmetric Anderson lattice is similar to that in the spin-$\frac{1}{2}$ Kondo lattice.
Should be notes also, when the hybridization is small the Hamiltonian (1) can be  mapped into the Kondo lattice model with the exchange imtegral
$J\simeq -\frac{2\Gamma U}{\epsilon_g(\epsilon_g+U)}$.

\begin{figure}[tp]
     \centering{\leavevmode}
\includegraphics[width=.7\linewidth]{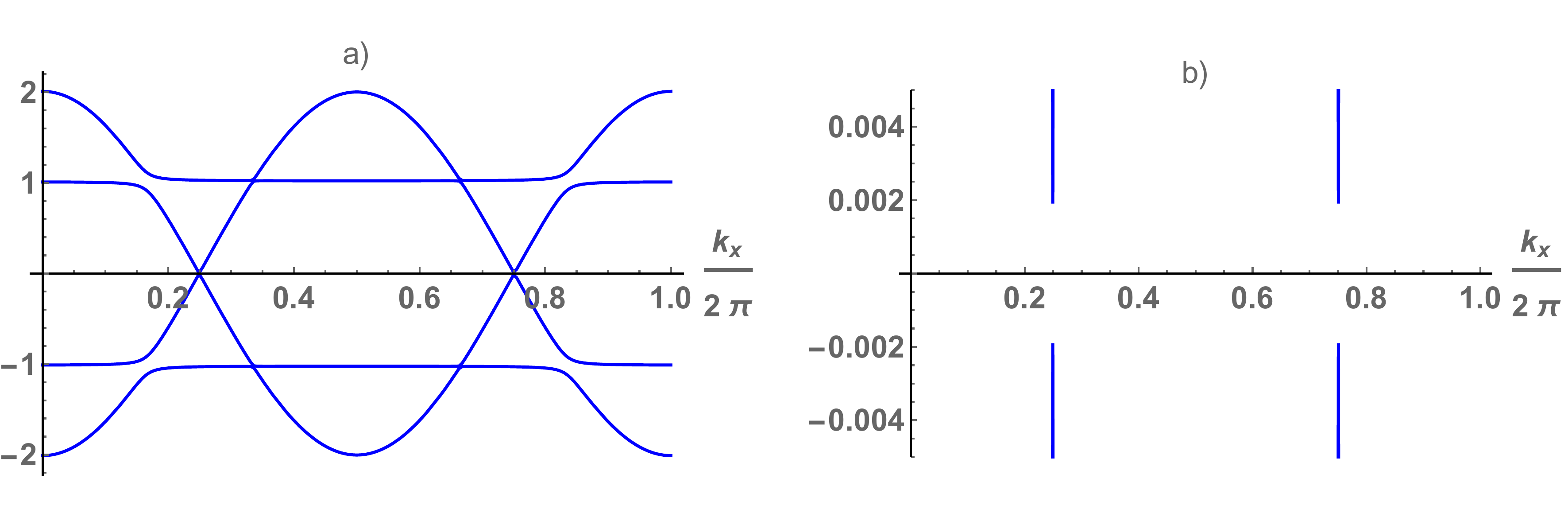}
\caption{(Color online)
Electron spectrum of the chain a) and low energy excitations b) (where $\Delta =0.004$) as function of wave vector calculated at $\epsilon_g=-1$ (or $U=2$), $v=0.1$, $\lambda=0.2$.
  }
\label{fig:1}
\end{figure}
\begin{figure}[tp]
\centering{\leavevmode}
\includegraphics[width=.7\linewidth]{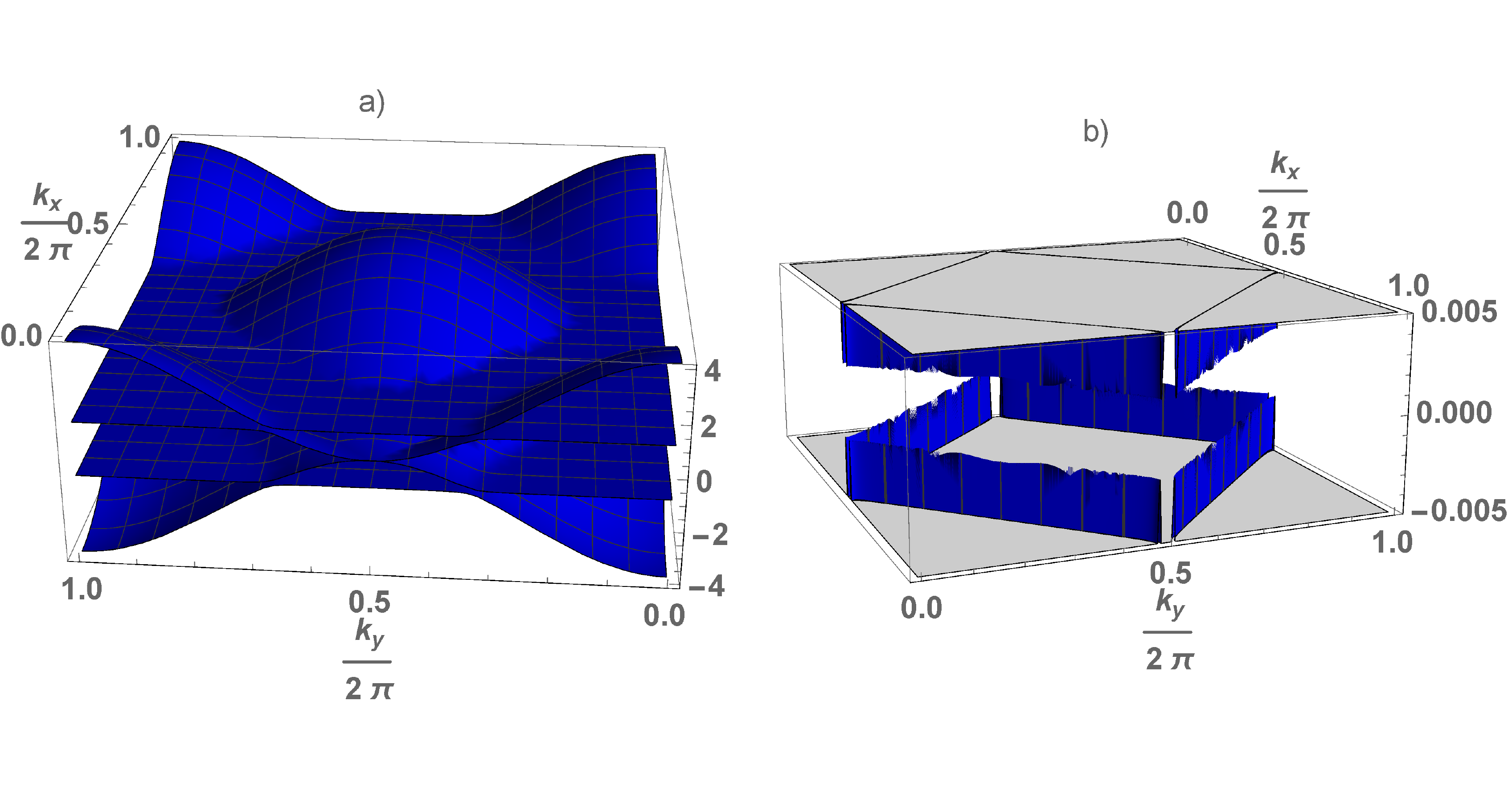}
\caption{(Color online)
Electron spectrum of the square lattice a) and low energy excitations b) (where $\Delta =0.004$) as function of the wave vector calculated at $\epsilon_g=-1$ (or $U=2$), $v=0.1$, $\lambda=0.2$.
  }
\label{fig:2}
\end{figure}
\begin{figure}[tp]
     \centering{\leavevmode}
\includegraphics[width=.5\linewidth]{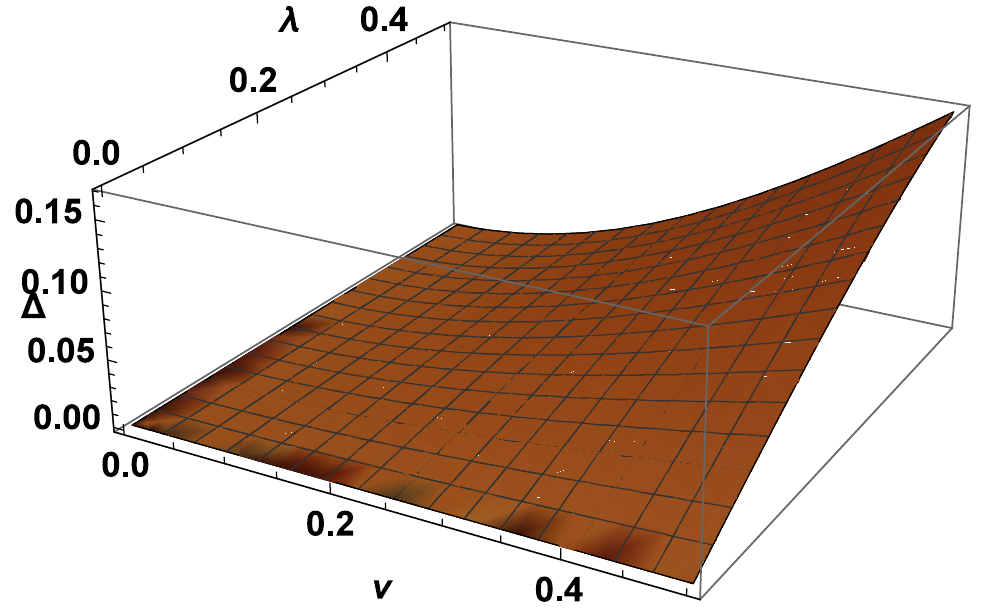}
\caption{(Color online)
The gap  $\Delta$  in the electron spectrum as function of $v$ and $\lambda$ calculated at $\epsilon_g=-1$ (or $U=2$), in the coordinates $v$, $\lambda$
(the value of $\Delta$ does not depend on the dimension of the lattice).
  }
\label{fig:3}
\end{figure}

\section*{The ground-state}
$\lambda$-field breaks the spontaneous symmetry \cite {K3,K4}, the magnetic moments of conduction and d-electrons are not conserved, the total number of electrons is conserved \cite{K1}. Only for $\textbf{q}=\overrightarrow{\pi}$, the electron spectrum corresponds to the symmetric Anderson lattice. Thus, we are talking about the lattice with the doubled period, as in the case of the Peierls phase-transition.
In this case the electron spectrum is symmetric with respect to zero and has the following form for the chain (see in Fig 1a) and square lattice (see in Fig 2a). Four branches of the spectrum $\pm E_\gamma (\textbf{k)} $ ($\gamma =1,2$) are determined by the following expression (see section Methods))
\begin{equation}
E_\gamma(\textbf{k}) = \frac{1}{\sqrt2} \sqrt{\alpha(\textbf{k})+(-1)^\gamma \sqrt{\alpha^2(\textbf{k})-4\beta^2(\textbf{k})}},
\label{eq:H5}
\end{equation}
where $\alpha (\textbf{k})= \varepsilon^2(\textbf{k})+ \epsilon_g^2 +\lambda^2 +2\Gamma$, $\beta^2 (\textbf{k})=(\lambda^2 +\epsilon_g^2)\varepsilon^2(\textbf{k})-2\epsilon_g \Gamma \varepsilon (\textbf{k})+\Gamma^2$.

For $v \neq 0$, $\lambda \neq 0$ and an arbitrary dimension of the lattice the spectrum is gapped at half-filling. The gap is equal to zero at $v=0$ or $\lambda=0$ (see in Fig 3), $v =\lambda =0$ corresponds to the atomic limit of the symmetric Anderson lattice with a bare on-site repulsion.

The dependence of the gap on $v$ and $\lambda$  has quite universal in nature, its value  does not depend on the dimension of the model (see in Fig 3). As shown below, the value of $\lambda$ is the solution of Eq (3), at given $U$ its value is determined by $v$, depends on dimension of the lattice.
In the chain the low energy spectrum is determined by two special points (see in Fig 1b). In the square lattice, they form a square in the $\textbf{k}$-plane (see in Fig 2b). The k-surface,  formed by the points on the $ \textbf {k} $-plane of the low-energy spectrum, is constructed as follows: each projection of the wave vector corresponds to two values of its other projections. The $\gamma$-branches of the spectrum are hybridized, the electron density is defined as $\sum_{\gamma=1,2}\int d\textbf{k} =n_s +1$, here $n_s$ is the density of s-electrons ($n_s=1$ at  half-filling). Thus, we can  talk about a large Fermi surface in the Kondo lattice also, the behavior of which is similar to the symmetric Anderson lattice.

In the ground state the equation for $\lambda$ follows from the form of the action (6) and the spectrum (2)
\begin{equation}
\frac{\lambda}{U}= \frac{1}{4}\sum_{\gamma=1,2}\int d \textbf{k} \frac{1}{E_\gamma(\textbf{k})} \frac {\partial E_\gamma^2(\textbf{k})}{\partial \lambda},
\label{eq:H6}
\end{equation}
where the value of the on-site repulsion corresponds to a condition $U=-2\epsilon_g$. The energy of the quasi-particle excitations depend on $\epsilon_g$ (2), thus (3) is self-consistent equation. As a result, the state of the symmetric Anderson lattice can be realized for  $\lambda \neq 0$.

As we noted above the gap in the spectrum depends both $v$  and $\lambda$ (in the Anderson model the scattering matrix depends on $v$ and $U$ \cite{W1}). Numerical solutions for $\lambda$ as function of $v$ are presented in Fig 4 for different dimension of the lattice. Results of calculations are obtained at $\epsilon_g=-1$, $U=2$ (see in Fig 4a)) and $\epsilon_g=-1/2$, $U=1$ (see in Fig 4b)).  Trivial solution $\lambda =0$ and $v=0$ corresponds to the symmetric Anderson model with $\epsilon_g=-\frac{U}{2}$ in which s- and d-electrons are not coupled. Nontrivial solution for $\lambda$ and $v$ determines  state of electron liquid in the symmetric Anderson model. Solution for $\lambda$ takes place  at finite values of $v$, the $\lambda$-value increases with increasing $v$ (see in Fig 4). Such at $v=0.45$ the gaps in the spectrum have the following values for U=2 (U=1): $\Delta =0.015$ ($\Delta =0.093$) in the chain, $\Delta= 0.018$  ($\Delta =0.074$) in square and $\Delta =0.005$  ($\Delta =0.088$) in cubic lattices. The value of the gap decreases with increasing $U$.

The stability of gapped ground state also depends on the magnitude of the on-site  repulsion. Comparing actions (6) at T = 0K for the gapped (paramagnetic) state and the phase state with the maximum magnetic moment $M=M_s+M_d=\frac{N}{2}+\frac{N}{2}$, we determine the region of stable insulator state. The minimum value of the  on-site Hubbard repulsion $ U_c $, at which the gapped state is realized, is $ U_c = 0.999205$ (1D),  $U_c = 0.99933$ (2D), $ U_c = 0.99914$ (3D) at $v=\frac{1}{2}$ and $ U_c = 0.96987$ (1D), $U_c = 0.98969$ (2D), $U_c = 0.97796 $ (3D) at $v=1$. The value of $ U_c $ weakly depends on the dimension of the lattice and the magnitude of the s-d hydridization $ v $. The non-trivial solution for $ \lambda $, shown in Fig. 4, corresponds to stable insulator state at $ U = 2 $, the gapped state at $ U = 1 $, is unstable.
Calculating the actions $ \delta S = \frac {S (\lambda)} {\beta} - \frac {S (0)} {\beta} $ at T = 0K,  which correspond to the states with the  gap (for $ \lambda \neq 0 $) and gapless with a maximum magnetic moment (for $ \lambda = 0 $), we determine the stability of the insulator state at $ U = 2 $. The numerical calculations are shown in Fig 5 for different dimensions of the lattice. The region of stability of the insulator state is highlighted in color, the solution for $ \lambda $ obtained for $ U = 2 $ (see Fig. 4a) is in the colored region. Note, that $\delta S=0$ at $\lambda =0$ and $\delta S \sim \lambda^2$ at $\lambda\to \infty$, thus intermediate values of $0<\lambda < 0.1$ correspond to the stable insulator state (see in Fig 5).

Taking into account the spectrum of quasi particle excitations (2) in  the action (6), we determine the low temperature behavior of the specific heat (its electronic part) for 3D system (cubic lattice). In the insulator state, the value of $ C $ exponentially depends on the value of the gap. Numerical calculations of the temperature dependence $\frac{C}{T}$ are shown in Fig. 6. The calculations are made for $ U = 2 $ and $ v = 0.43$, $ \lambda =  0.0075$, $ \Delta = 0.0023$ (see Fig. 6a) and $ v = 0.5$, $ \lambda = 0.07 $, $ \Delta = 0.0279 $ (see Fig.6b). Note that for these parameters, the gaps in the electron spectrum differ by an order of magnitude.
The experimentally measured heat capacity $ \frac {C} {T} $ $ SmB_6$ \cite {C} shows the anomalous behavior in the low temperature range $ 2K <T <10K $, namely at T = 4K there is a minimum. According to numerical calculations (see insets in Fig. 6), a local minimum  of $ \frac {C} {T} $ is realized at low temperatures. The temperature range and the minimum value depend on the value of the gap.

\begin{figure}[tp]
     \centering{\leavevmode}
\begin{minipage}[h]{.75\linewidth}
\center{
\includegraphics[width=\linewidth]{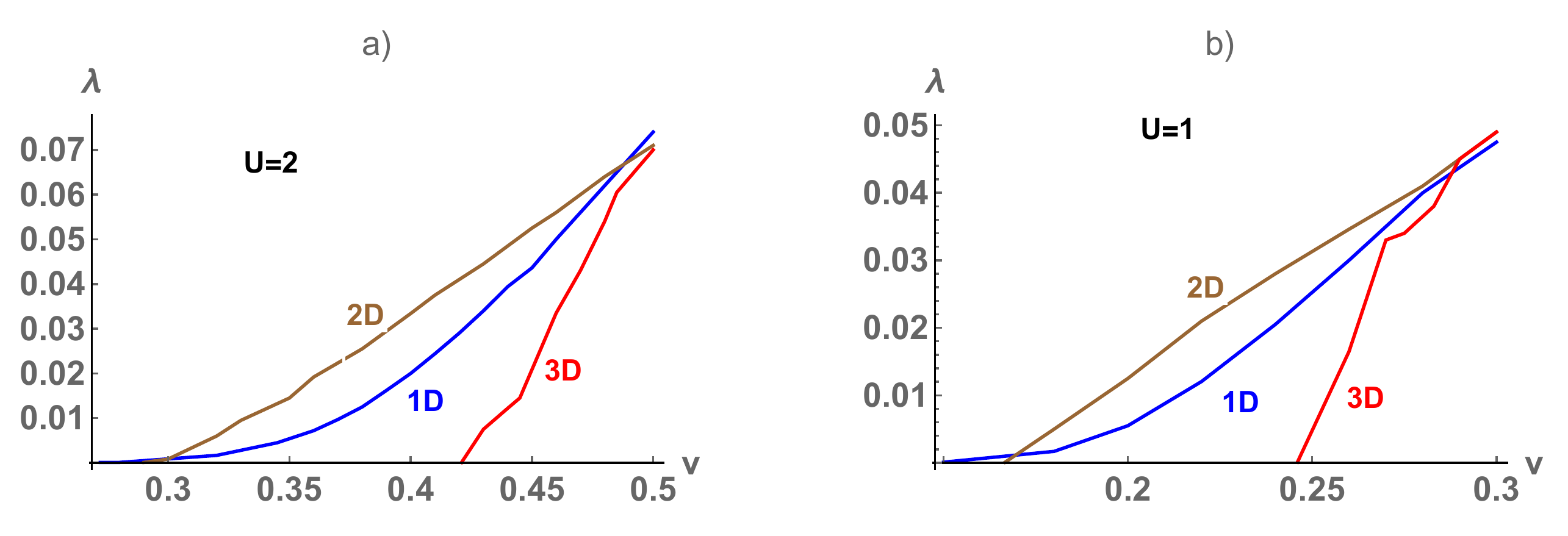}
                 }
   \end{minipage}
\caption{(Color online)
$\lambda$ as a function of $v$ calculated at $\epsilon_g=-1$, $U=2$  a) and  $\epsilon_g=-0.5$, $U=1$ b) for the chain, square and cubic lattices.
  }
\label{fig:4}
\end{figure}
\begin{figure}[tp]
\includegraphics[width=\linewidth]{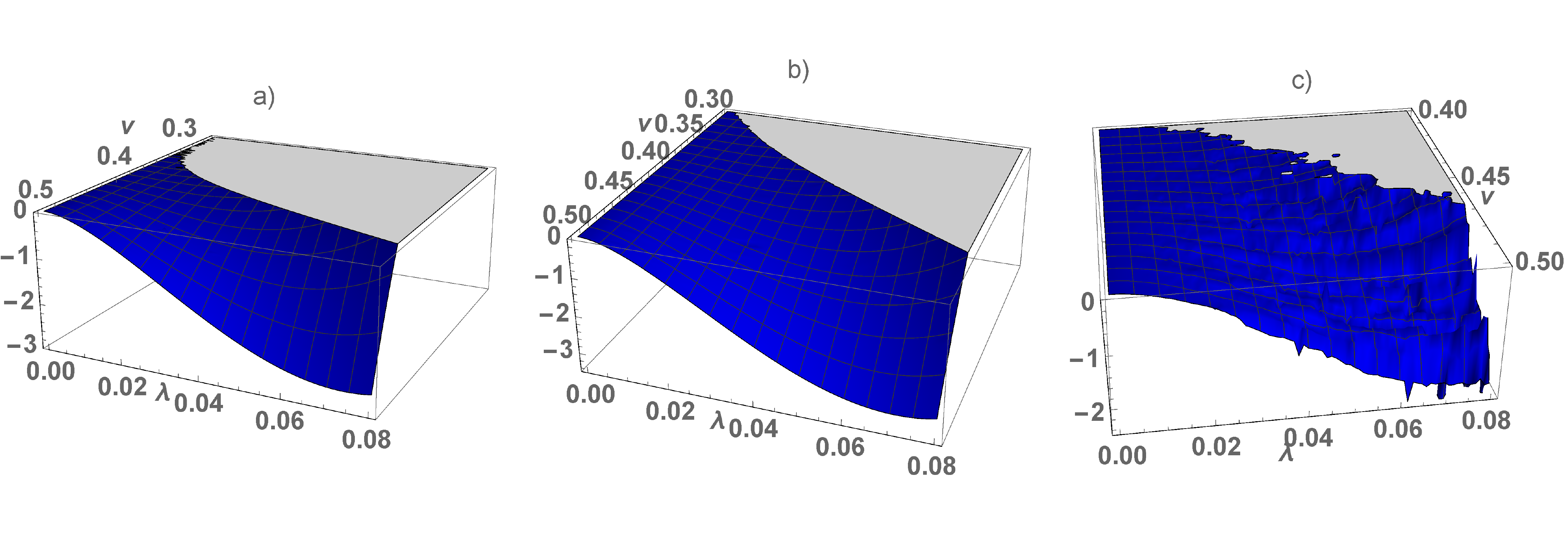}
\caption{(Color online)
$\delta S \times 10^5 $ as a function of $v$ and $\lambda$  calculated at $\epsilon_g=-1$ ($U=2$) for the chain a), square b) and cubic c) lattices,
the region of stability of the gapped state is highlighted in color.
  }
\label{fig:5}
\end{figure}
\begin{figure}[tp]
   \centering{\leavevmode}
\begin{minipage}[h]{.9\linewidth}
\center{
\includegraphics[width=\linewidth]{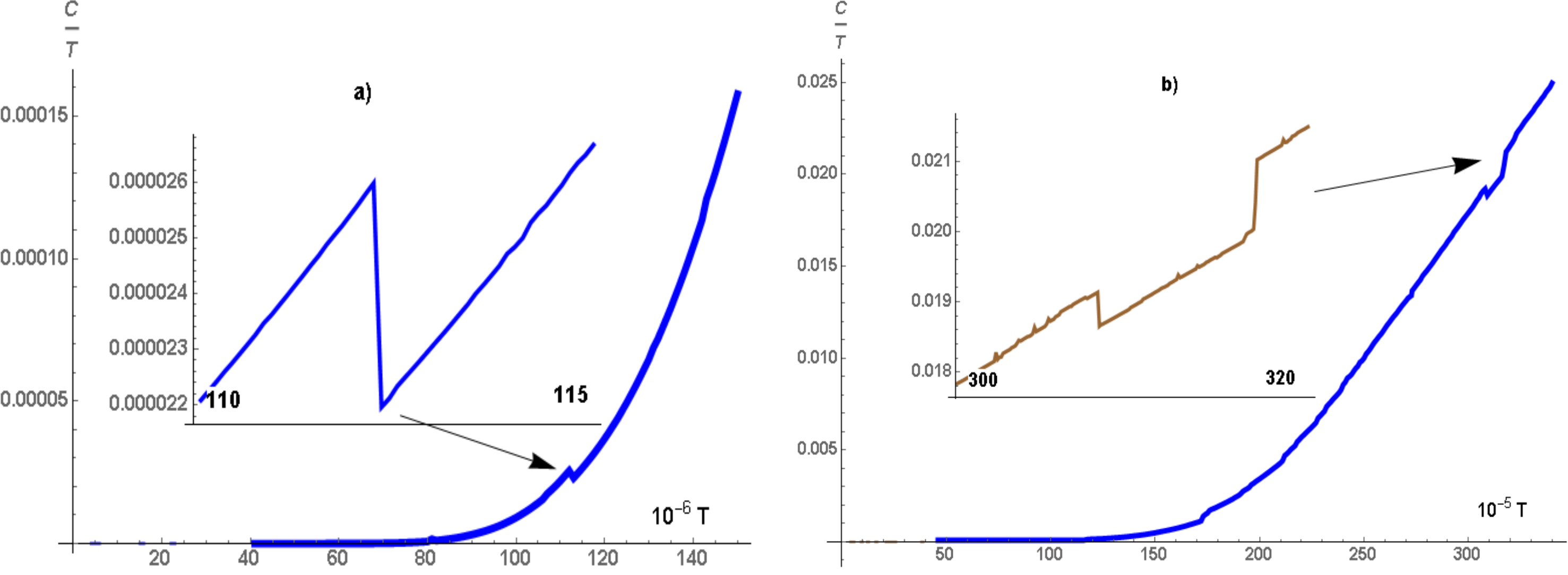}
                 }
   \end{minipage}
\caption{(Color online)
The temperature dependence of the specific heat $\frac{C}{T}$ calculated for cubic lattice at $U=2$ : $ v = 0.43$, $ \lambda =  0.0075$, $ \Delta = 0.0023$ a) and $ v = 0.5$, $ \lambda = 0.07 $, $ \Delta = 0.0279 $. Inserts illustrate the regions of  abnormal behavior of the specific heat at low temperatures.
  }
\label{fig:5}
\end{figure}

\section*{Methods}
\subsection*{Canonical functional of the symmetric Anderson model}

Let us introduce the operator $\chi_{j}^\dagger= d^\dagger_{j, \uparrow}d_{j,\downarrow}$ and redefine the term ${\cal H}_{int}$ (1) is the following form
${\cal H}_{int}= -U\sum_{j}\chi_{j}^\dagger\chi_{j}$ \cite{K1}.
The Hubbard-Stratonovich transformation converts the interacting problem into a non-interacting one in a stochastic external field (hereinafter we will define it as the $ \lambda $-field). We define the interaction term, taking into account the action ${S}_{0}$
\begin{eqnarray}
&& S=S_{0}+\sum_{j} \frac{\lambda^\ast_{j}\lambda_{j}}{U}+\sum_{j}(\lambda_{j} \chi_{j}+ H.c.)
 \label{A1}
\end{eqnarray}
The canonical functional is determined as $${\cal Z}=\int {\cal D}[\lambda] \int {\cal D}[\chi\dagger,\chi] e^{-S},$$ where the action $S=\frac{1}{U}\sum_{j}\lambda^\ast_{j}\lambda_{j}+
\int_0^\beta d\tau \sum_\textbf{k}\Psi_\textbf{k}^\dagger (\tau)[\partial_\tau  + {\cal H}_{eff}(\textbf{k})]\Psi_\textbf{k} (\tau)$,  $\Psi_\textbf{k} (\tau)$ is  the wave function, $\textbf{k}$ is the wave vector of an electron. We expect that $ \lambda_{j} $ does not depend on $ \tau $, since the translation invariance is conserved in an electron liquid state.

At the on-site hybridization between d-states of electrons with different spins and due to translation invariance of the model Hamiltonian, only the phases of $\lambda_j$  depends on $j$, a namely $\lambda_{j}=\exp(i \textbf{q }\textbf{j}) \lambda$ \cite{K1}, where $\textbf{q}$ is an unknown wave vector.
The phase can also fluctuate at the lattice site; subsequent averaging over local fluctuations restores translational invariance \cite{D}.
The task is reduced to moving fermions in a static inhomogeneous $\lambda$-field that sets the form of ${\cal H}_{eff}(\textbf{k})$
\begin{equation}
{\cal H}_{eff}(\textbf{k}) = \left(
\begin{array}{cccc}
-\varepsilon (\textbf{k}) & v & 0 & 0\\
v & \epsilon_g & \lambda & 0\\
0 & \lambda^*& - \epsilon_g& v \\
0 & 0& v & -\varepsilon (\textbf{k}+\textbf{q})
\end{array}
\right)
\end{equation}
The spectrum of non-interacting s-electrons is given by $\varepsilon (\textbf{k})=-2\sum_{i=1}^D \cos k_i$, here D is dimension of the model.

We can  integrate out fermions to obtain  the following action $S$ per an atom
\begin{eqnarray}
\frac{S(\lambda)}{\beta}=-\frac{{T}}{{N}}\sum_{\textbf{k}}\sum_n \sum_{\gamma=1}^{4} \ln [-i \omega_n+E_\gamma(\textbf{k},\textbf{q})]
+\frac{|\lambda|^2}{{U}},
 \label{A2}
\end{eqnarray}
where $\omega_n =T(2n+1)\pi$ are the Matsubara frequencies, $\textbf{k}$, $\textbf{q}$ are the momenta of electrons,  four quasiparticle excitations $E_\gamma(\textbf{k},\textbf{q})$ ($\gamma =1,...,4$) determine the electron states in the  $\lambda$-field. In the saddle point approximation the canonical functional ${\cal Z}$ will be dominated by the minimal action $S$ (6), that satisfies the following equation $\partial S/\partial \lambda =0$.
In the Kondo problem, the processes of electron scattering from spin flip dominate, in our case, the $\lambda$-field connects the states of s-electrons with opposite spins and different momenta $\textbf{k}$ and $\textbf{k}+\textbf{q}$.  Due to the on-site repulsion between electrons in the Hubbard model\cite{K1}, an effective field connects states of electrons with different spins and different momenta $\textbf{k}$, $\textbf{k}+\overrightarrow{\pi}$. This state is stable at half-filling, a gap opens in the electron spectrum \cite{K1}. In this case, the same mechanism of coupling between  s-electrons is realized indirectly through an intermediate subsystem of d-electrons. The gap in the electron spectrum will naturally be smaller.

\section*{Conclusion}

Using mean field approximation we have considered the solution of the symmetric Anderson lattice at half-filling for different dimensions of the lattice.
It is shown that an effective field, which binds the states of s-electrons with different spins and momenta, leads to the gap in the electron spectrum. The electron spectrum is mirror  symmetric (with respect to zero energy), has the type of the Majorana spectrum. The states of s- and d-electrons are hybridized, therefore the Fermi surface is determined by the total density of electrons. In the symmetric Anderson lattice model the gaped state of electron liquid corresponds to breaking spontaneous symmetry.
This made it possible to take into account the processes of spin flip scattering of conduction electrons on localized electrons. This is important because it is these processes that lead to the appearance of the  Abrikosov - Suhl resonance in the Kondo problem and an insulator state in the Kondo lattice.
The gapped state is formed at  finite values of the s-d hybridization and on-site repulsion.
Numerical calculations of the low-temperature behavior of the heat capacity are carried out in the gapped state.
The results explained the presence of a local minimum of the heat capacity $ \frac {C} {T} $ at low temperatures observed in $SMB_6$.
In the symmetric Anderson lattice, a local spin-$\frac{1}{2}$ is realized at lattice site,  therefore its behavior is similar to the spin-$\frac{1}{2}$ Kondo lattice. The proposed approach allows us to describe the symmetric Anderson and Kondo lattices in one formalism. One can speak of a large Fermi surface and the gapped state of an electron liquid in the Kondo lattice at half filling.

\section*{Author contributions statement}
I.K. is an author of the manuscript
\section*{Additional information}
The author declares no competing financial interests.

\end{document}